\documentclass[aps,prd,amssymb,showpacs,floatfix,twocolumn,superscriptaddress,nofootinbib]{revtex4-1}
\usepackage{amsmath,amsfonts,graphics,epsfig,amssymb,bm}
\usepackage{multirow}
\usepackage{enumerate}
\usepackage{graphicx}
\usepackage [latin1]{inputenc}
\usepackage[colorlinks, citecolor=blue, anchorcolor=blue, linkcolor=blue]{hyperref}

\usepackage[doipre={doi:~}]{uri}

\begin{document}
%\captionsetup[figure]{labelformat={default},labelsep=period,name={FIG.}}

\title{Revisiting the constraints on primordial black hole abundance with the isotropic gamma ray background}
\author{Siyu Chen}
\author{Hong-Hao Zhang}
\email[Corresponding author. ]{zhh98@mail.sysu.edu.cn}
\author{Guangbo Long}
\email[Corresponding author. ]{longgb@mail2.sysu.edu.cn}
\affiliation{School of Physics, Sun Yat-sen University, Guangzhou, GuangDong, People's Republic of China}

\begin{center}
\begin{abstract}
 We revisit the constraints on primordial black holes (PBHs) in the mass range $10^{13}-10^{18}$\,g by comparing the 100\,keV-5\,GeV gamma-ray background with isotropic flux from PBH Hawking radiation (HR). We investigate three effects that may update the constraints on the PBH abundance; i) reliably calculating the secondary spectra of HR for energy below 5\,GeV, ii) the contributions to the measured isotropic flux from the Galactic PBH HR and that from annihilation radiation due to evaporated positrons, iii) inclusion of astrophysical background from gamma-ray sources.
  The conservative constraint is significantly improved by more than an order of magnitude at $2\times10^{16}$\,g\,$\lesssim M\lesssim 10^{17}\,$g over the past relevant work, where the effect ii is dominant. After further accounting for the astrophysical background, more than a tenfold improvement extends to a much wider mass range $10^{15}$\,g\,$\lesssim M\lesssim 2\times 10^{17}\,$g.
  %such improvements on the flux analysis lead to,  but have been neglected by relevant study %

\end{abstract}
\end{center}
\maketitle

\section{Introduction}
\label{sec:intro}
Primordial black holes (PBHs) are the only candidate that can solve the dark matter (DM) problem without involving new physics~\cite{Carr2020,Khlopov2010}. At present, there are still open PBH mass windows (e.g.,\,$10^{17}$-$10^{22}$\,g) that can constitute all or most of the DM all over the universe, see e.g.,\,\cite{Carr2020}. If the PBHs contribute significantly to the DM, then the time-integrated Hawking radiation (HR) of PBHs with a mass of about $10^{13}$-$10^{18}$\,g should significantly affect on the observed isotropic diffuse gamma-ray background (IGRB) and/or cosmic X-ray background (CXB) for energy above 100\,keV~\cite{Carr:2009jm,Arbey:2019vqx,Ballesteros:2019exr,Iguaz:2021irx,Carr2020,Auffinger2022}. In order for HR to propagate through the intergalactic medium, these PBHs must survive at least into the transparent age of the cosmic microwave background (CMB)~\cite{Arbey:2019vqx}. This condition sets the above PBH mass lower limit.
In addition, the energy at the peak of HR decreases with increasing PBH mass. The contribution from PBHs larger than $10^{18}$\,g is concentrated in CXB below 100\,keV, but is negligible~\cite{Arbey:2019vqx}.

Carr\,{\it et\,al.}~\cite{Carr:2009jm} limited the PBH abundance conservatively within about $10^{13}$-$10^{18}$\,g based on the evaporated contribution of PBHs that did not exceed the observed extragalactic photon background in the 100\,keV-100\,GeV range. Recently, Arber\,{\it et\,al.}~\cite{Arbey:2019vqx} and Carr\,{\it et\,al.}~\cite{Carr2021} updated the bound with updated background observations (new Fermi-LAT data), and the former~\cite{Arbey:2019vqx} also studied extended mass functions of Kerr black holes.
 Ballesteros\,{\it et\,al.}~\cite{Ballesteros:2019exr} and Iguaz~{\it et\,al.}~\cite{Iguaz:2021irx} set tighter bound in the mass range $10^{16}$-$10^{18}$\,g with hard CXB and soft IGRB taking into account the emission from AGNs (Active galactic nucleus), respectively. It is worth emphasizing that~\cite{Iguaz:2021irx} considers the contribution of the Galactic PBH emission and the annihilation radiation due to evaporated positrons to the measured isotropic flux.

In addition to the above constraints using the isotropic cosmic photon background, there are many other limits on PBHs in the mass range $5\times10^{14}$-$10^{18}$\,g based on the inconsistency between predicted HR-induced signatures and actual observations. These observations can be electron or positron cosmic rays~\cite{Boudaud2019}, the cosmic microwave background~\cite{Poulin:2016anj,Stcker2018,Clark2017}, gamma-rays and X-ray fluxes from specific objects~\cite{DeRocco:2019fjq,Laha:2019ssq,Laha:2020ivk,Laha:2020vhg}, neutrinos~\cite{Wang:2020uvi,Dasgupta:2019cae}, and so on. Their resulting excluded region of PBH mass and abundance are about similar to those from the cosmic photon background.

In this work, we aim to improve the constraints on the PBH abundance by comparing the simulated isotropic flux of PBHs and astrophysical sources with the observed IGRB in the {\bf0.1\,MeV-5\,GeV} range~\footnote{In general, IGRB is referred as obervations of~\cite{Fermi-LAT:2014ryh} and~\cite{EGRET:1997}. For convenience, all data used in this paper is broadly referred to as IGRB.}. Compared with past relevant works, in particular Ref.\,\cite{Arbey:2019vqx}, three improvements on the treatments of the isotropic flux that may eventually improve the constraints will be investigated: (i) The IGRB from unresolved sources, such as AGNs and SFGs (star-forming galaxies), is modeled according to~\cite{Roth:2021lvk} and \cite{Murase:2019vdl}. (ii) We reliably model the secondary spectrum of PBH HR below 5\,GeV with the latest public code \texttt{BlackHawk}.\,2.0~\cite{Arbey:2021mbl}, which calculates the final state
radiation (FSR) and decays of PBHs's primary particles (leptons and pions) below 5\,GeV as discussed in~\cite{Coogan:2020tuf}. (iii) We extend the analysis of Ref.\,\cite{Iguaz:2021irx} from the observed 10\,keV-10\,MeV flux to our 100\,keV-5\,GeV one. We take into account the diffuse flux contributed by the Galactic, which is measured by the detectors but cannot be separated from the truly extragalactic contribution, and an indirect component by the $e^{\pm}$ annihilation via a positronium generated by the charge exchange between
 atomic hydrogen and positron evaporated from PBHs~\cite{Iguaz:2021irx}.

The plan of this paper is as follows.
In Sec.~\ref{secII} and~\ref{secIII}, we provide computational details of the isotropic flux from PBH HR, illustrate the data sets of cosmic background spectrum, and describe the models for the astrophysical background.
Then, in Sec.~\ref{sec:result} we analyze the data and give our new constraints on the PBHs abundance. Finally, discussions and conclusions are presented in Sec.~\ref{sec:discussion} and~\ref{sec:conclusion}.

\section{Hawking radiation from PBHs}
\label{secII}
\subsection{Hawking radiation from singe PBH}
In this work, we assume the mass distribution of PBHs is monochromatic and let us denote $E$ as the photon energy in the local cosmic frame. The total photon spectrum $\frac{d{\dot{N}}}{d{E}}$ emitted by a single PBH with mass $M$, in unit energy and unit time, can be written as (see e.g.,~\cite{Carr:2009jm})
\begin{equation}
    \frac{d{\dot{N}}}{d{E}}=\frac{d{\dot{N}^{\rm pri}}}{d{E}}
    +\frac{d{\dot{N}^{\rm sec}}}{d{E}}.
    \label{Fpbh}
\end{equation}
The primary component $\frac{d{\dot{N}^{\rm pri}}}{d{E}}$ results from the direct Hawking emission, which is similar to the blackbody radiation but with a greybody factor counting the probability that a Hawking particle evades the PBH gravitational well. The secondary component (the second term) comes from the decay of the hadrons produced by the fragmentation of primary quarks and gluons.

 In recent years, the HR spectra $\frac{d{\dot{N}}}{dE}$ are usually calculated with the popular public code \texttt{BlackHawk}.\,1, e.g.,~\cite{Arbey:2019vqx,Ballesteros:2019exr,Iguaz:2021irx}. However, for the primary particles with energy below 5\,GeV, it uses ``extrapolation tables" to compute the secondary spectra, and thus these spectra are unreliable~\cite{Coogan:2020tuf}. Thanks to Coogan\,{\it\,et\,al}.'s work~\cite{Coogan:2020tuf}, the latest version of this code (\texttt{BlackHawk}.\,2.0~\cite{Arbey:2021mbl}) incorporating the method of~\cite{Coogan:2020tuf} can reliably simulate the secondary spectra from the FSR and the decay of the primary particles. This method takes advantage of the Altarelli-Parisi splitting functions to model the FSR and uses new \texttt{Hazma} to compute the photon spectrum from decays by~\cite{Coogan:2020tuf}:
\begin{align}
    &\frac{d{\dot{N}^{\rm sec}}}{dE} = \sum_{i=e^{\pm},\mu^{\pm},\pi^{\pm}}\int d{E_{i}}
    \frac{d\dot{N}^{\rm pri}_{i}}{dE_{i}}
    \frac{dN^{\rm FSR}_{i}}{dE}\notag\\
    &\quad + \sum_{i=\mu^{\pm},\pi^{0},\pi^{\pm}}\int d{E_{i}}
    \frac{d\dot{N}^{\rm pri}_{i}}{dE_{i}}
    \frac{dN^{\rm decay}_{i}}{dE},
    \label{FSR}
\end{align}
where $\frac{d\dot{N}^{\rm pri}_{i}}{dE_{i}}$ are the primary spectra of leptons and pions. The explicit expression of $\frac{dN^{\mathrm{decay}}}{dE}$ and the FSR spectra $\frac{dN^{\rm FSR}}{dE}$ can be seen in~\cite{Coogan:2019qpu} and~\cite{Coogan:2020tuf}, respectively.

\begin{figure}[htbp]
  \centering
    \includegraphics[width=0.45\textwidth]{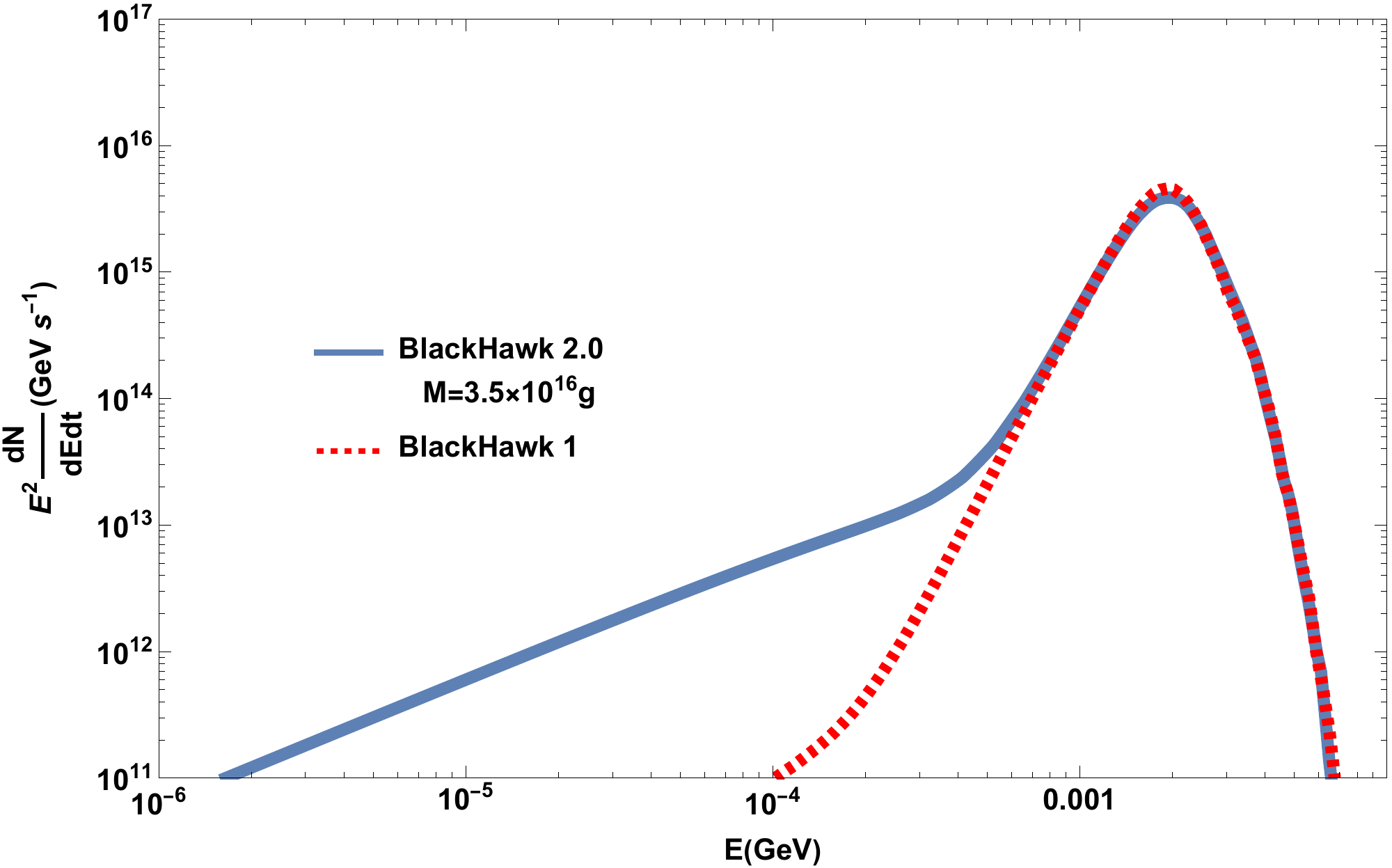}
    \includegraphics[width=0.45\textwidth]{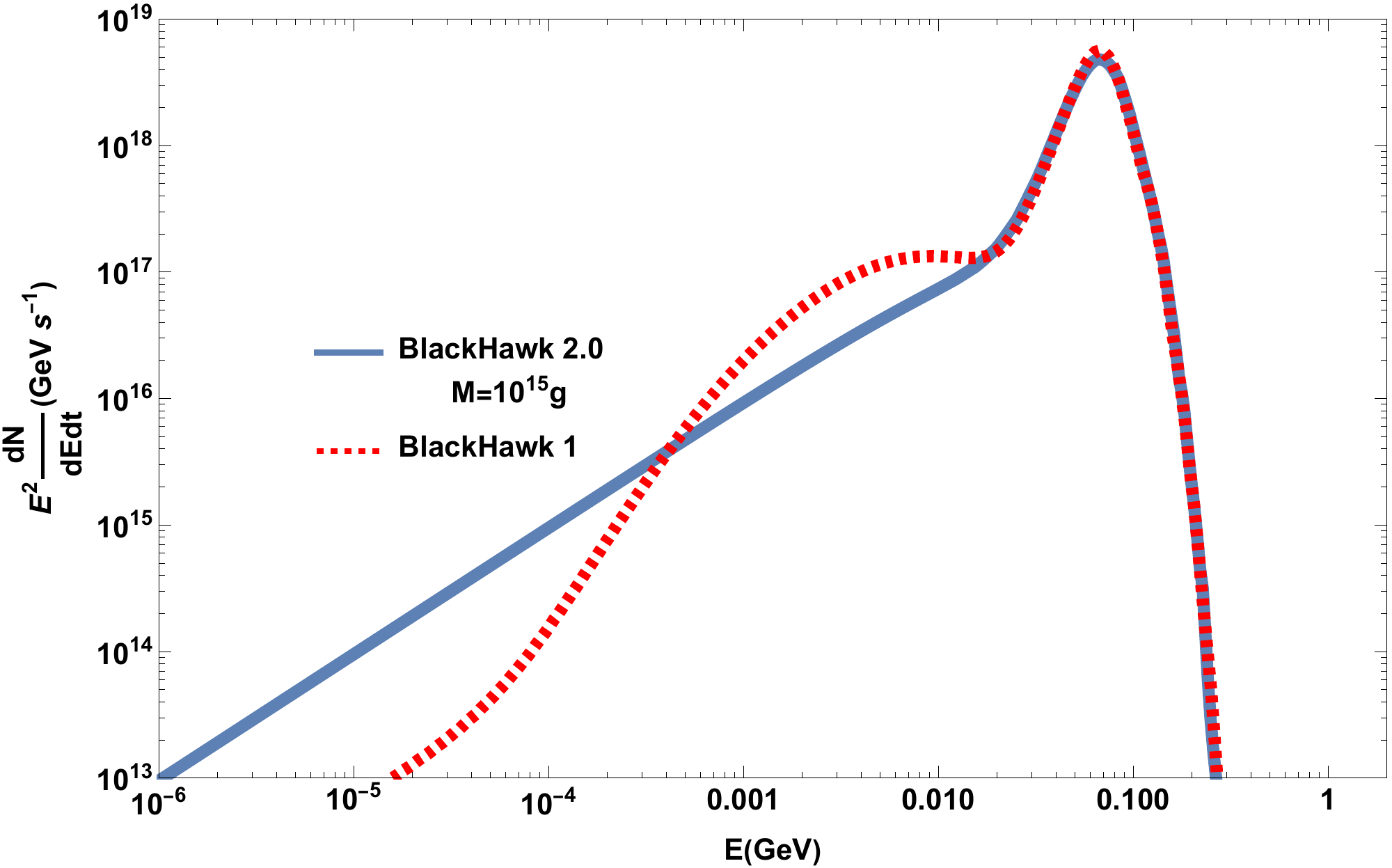}
    \caption{The secondary spectra for $M=3.5\times10^{16}$\,g (top) and $M=10^{15}$\,g (bottom) obtained by different version of \texttt{BlackHawk}. The solid line represents that the spectrum is calculated with Eq.\,(\ref{FSR}) using \texttt{BlackHawk}.\,2.0~\cite{Arbey:2021mbl} and the dash line represents one from \texttt{BlackHawk}.\,1.2~\cite{Arbey2019BPH}.}
    \label{fig:FSR}
\end{figure}

As an improvement over the past relevant works \cite{Arbey:2019vqx,Ballesteros:2019exr,Iguaz:2021irx}, we simulate the secondary photon spectrum and the primary spectra of all particles with \texttt{BlackHawk}.\,2.0~\cite{Arbey:2021mbl}. Fig.\,\ref{fig:FSR} shows the resulting low energy correction in the secondary spectrum. For small PBH mass range, e.g.,\,$M=10^{15}$\,g, the new method (the blue line in Fig.\,\ref{fig:FSR}) gives higher flux in the lower energy band than \texttt{BlackHawk}.\,1. For large mass, e.g.,\,$M=3.5\times 10^{16}$\,g, it first gives weaker flux and then turns to higher in the lower energy band.

\subsection{The flux from extragalactic and Galactic PBHs}
 We consider the total diffuse flux emitted by PBHs throughout the universe. The flux can be separated to an extragalactic and Galactic part as
\begin{equation}
    \frac{dN}{dE}=\frac{dN_{\rm EG}}{dE}+\frac{dN_{\rm Gal}}{dE}.
    \label{Ftot}
\end{equation}
 Each part includes a direct photon component and an indirect one from $e^+$ annihilations~\cite{Iguaz:2021irx}, corresponding to the photon spectrum $\frac{d{\dot{N}}}{dE}$ and $\frac{d{\dot{N}}^{\rm ann.}}{dE}$ from a single PBH respectively.

The measured extragalactic flux from a population of extragalactic PBHs of mass $M$ with different ages is the redshifted sum over all epoch emissions~\cite{Carr:2009jm,Arbey:2019vqx}
\begin{align}
\frac{dN_{\rm EG}}{dE}\approx\label{eq:FEG}
&\frac{c}{4\pi}\,n_{\rm PBH}(t_0)\,\int^{\rm min(t_0,\tau)}_{t_{\rm min}} d{t}\, \big(1+z(t)\big) \\ \nonumber
 &\Bigg(\frac{d{\dot{N}}_{\rm EG}}{dE}\Big(M(t),\big(1+z(t)\big)E\Big)+\\  \nonumber
 &\frac{d{\dot{N}}^{\rm ann.}_{\rm EG}}{dE}\Big(M(t),\big(1+z(t)\big)E\Big)\Bigg),\  \nonumber
\end{align}
where $M(t)$ is a time dependent mass of a PBH, $z(t)=(H_0t)^{-2/3}-1$ is the redshift parameter
with the Hubble constant $H_0$, and $n_{\rm PBH}(t_0)\approx\frac{f \rho}{M}$ represents the number density of PBHs for a given initial mass $M$ at the present universe's age $t_0$. Factor $f$=$\Omega_{\rm PBH}$/$\Omega_{\rm DM}$ is a fraction of the total DM density in the Universe, and $\rho=2.17\times 10^{-30}$ $\text{g\,cm}^{-3}$ denotes the current DM density of the Universe~\cite{Planck:2018vyg}. The integration runs from the time at last scattering of the CMB $t_{\rm min}$=$380\,000\,$years to $t_{\rm max} = {\rm Max}(\tau(M),t_0)$ where $\tau(M)\propto M^3$ is the lifetime of PBHs.

We assume the positrons emitted in the PBH evaporation can form positronium (Ps) with electrons of atoms in cosmological medium, following Ref.\,\cite{Iguaz:2021irx}. Since the case of $e^{\pm}$ annihilation via Ps formation ($f_{\rm P_{s}}$=1) is more realistic than the direct $e^{\pm}$ annihilation ($f_{\rm P_{s}}$=0), we only consider the former~\cite{Guessoum:2005cb}. An annihilation of para-positronium can yield two photons and an ortho-positronium one yields three, with a total energy of 2\,$\rm m_{e}c^2$~\cite{Prantzos:2010wi}. Thus, the indirect photon component in Eq.\,(\ref{eq:FEG}) can be written as
\begin{align}
\begin{split}
  \frac{d{N}^{\rm ann.}_{\rm EG}}{dE}&= \frac{c}{4\pi}\,n_{\rm PBH}(t_0)
    \Big({\dot N}_{\rm e^{+}}\big(M(t_{\rm c})\big)\cdot
    2\times\frac{1}{4}\,\frac{\rm m_{\rm e}c^2}{E}\\
    &\delta[{\rm m_{e}c^2}-E\big(1+z(t_{\rm c})\big)]
   +3\times\frac{3}{4}\,\int^{t_{\rm max}}_{t_{\rm min}}\, d{t}\,(1+z)\\
   &{\dot N}_{\rm e^{+}}\big(M(t)\big)\,\frac{1}{N_{3\gamma}}\frac{d N_{3\gamma}}{d[(1+z)E]}\Big)\,,
\end{split}
\label{annEG}
\end{align}
where ${\dot N}_{\rm e^{+}}=\int dE^{'}\frac{d{\dot N}_{\rm e^{+}}}{dE^{'}}$ is the number of $e^{+}$ emitted by a PBH in unit time~\cite{Iguaz:2021irx} and ${\dot N}_{\rm e^{+}}\big(M(t_{\rm c})\big)$ represents the $e^{+}$ number at $t_{\rm c}$, where $t_{\rm c}$ satisfies the equation $E\big(1+z(t_{\rm c})\big)=\rm m_ec^2$ and inequation $t_{\rm min}\leq t_{\rm c}\leq t_{\rm max}$. The factors $\frac{1}{4}$ and $\frac{3}{4}$ describe the rate of the emitted photons in para-positronium and ortho-positronium state, respectively~\cite{Prantzos:2010wi}. The energy spectrum of para-positronium annihilation is described by  the Heaviside step-function $\theta$ (it should be Dirac function if the photons are not redshifted). The normalized spectrum of ortho-positronium annihilation is denoted by~\cite{Murphy2005}
\begin{align}
\begin{split}
  &\frac{1}{N_{3\gamma}}\frac{d N_{3\gamma}}{d[(1+z)E]}= \frac{2}{(1+z)\,(\pi^2-9)\,\rm m_{e}c^2}\,\Big(\frac{1-x}{x\,(\frac{2}{x}-1)^2}+\\
  &\frac{2\,(1-x)}{x^2}\,\rm ln(1-x)-\frac{2\,(1-x)^2}{(2-x)^3}\,\rm ln(1-x)+\frac{2}{x}-1\Big)\,,
\end{split}
\label{f3}
\end{align}
where $x=E/\rm m_{e}c^2$ and the relation $\frac{1}{N_{3\gamma}}\frac{d N_{3\gamma}}{d[(1+z)E]}\,d[(1+z)E]=\frac{1}{N_{3\gamma}}\frac{d N_{3\gamma}}{dE}\,dE$ has been used.

\begin{figure}[htbp]
  \centering
    \includegraphics[width=0.45\textwidth]{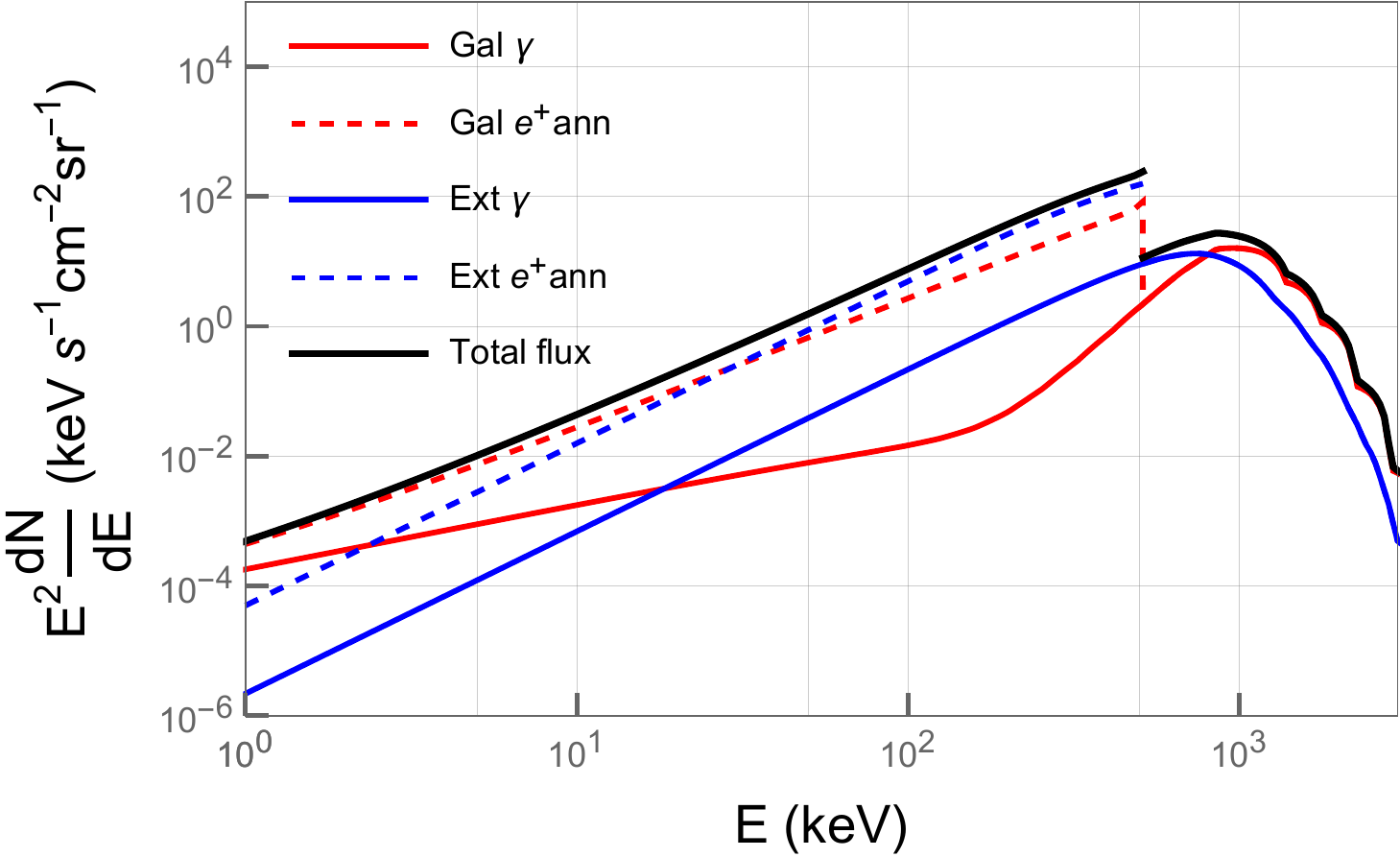}
    \includegraphics[width=0.45\textwidth]{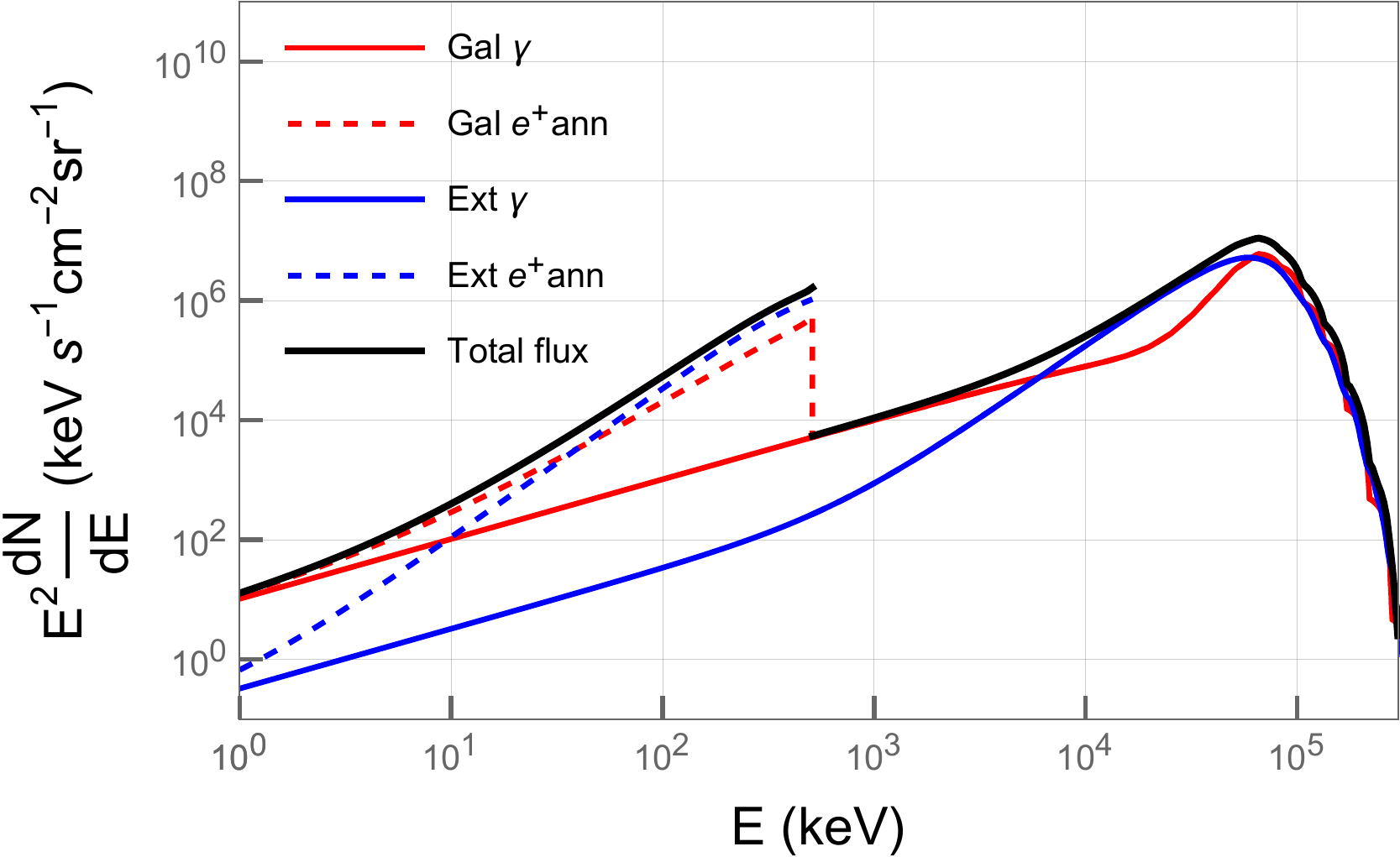}
    \caption{Different parts of the total diffuse
flux in Eq.\,(\ref{Ftot}) for $M=7\times 10^{16}$g, $f=1$ (top panel) and $M=10^{15}$g, $f=1$ (bottom panel): the direct Galactic and extragalactic contributions, and their indirect $e^{+}$ annihilation ones.}
    \label{fig:EGGal}
\end{figure}

Since we are embedded inside the Galactic halo, an diffuse flux from the Galactic PBHs cannot be separated into the truly extragalactic contribution when the flux is measured by detectors; see~\cite{Iguaz:2021irx} for details. Therefore, it should be taken into account in our simulation. As this flux depends on
the integral along the line of sight of the Galactic DM distribution, and a conservative estimation of it is given by~\cite{Iguaz:2021irx}
\begin{align}
	\frac{dN_{\rm Gal}}{dE}=\frac{f}{4\pi M}\,{\cal D}_{\rm min}\,( \frac{d{\dot{N}}_{\rm Gal}}{dE} +\frac{d{\dot{N}}^{\rm ann.}_{\rm Gal}}{dE})\,,
\label{EqGal}
\end{align}
where ${\cal D}_{\rm min}$ is the minimum of {\it D-factor}:
\begin{equation}
{\cal D}_{\rm min}\equiv \int_{\rm GAC} ds\, \rho_{g}\,,
\end{equation}
where GAC denotes the line of sight towards the Galactic anti-center, namely $l=180^{\circ}$ and $b=0^{\circ}$ in galactic coordinates, and $\rho_g$ is the DM distribution assumed a Navarro-Frenk-White profile with parameters $r_{s}=9.98$\,kpc, $\rho_{s}=2.2 \times 10^{-24}$ $\rm g\,cm^{-3}$~\cite{Karukes2020}.

The photon spectrum of positronium annihilation is
\begin{align}
\begin{split}
  \frac{d{\dot{N}}^{\rm ann.}_{\rm Gal}}{dE}&=\dot N_{\rm e^{+}}\,\big(
    2\times\frac{1}{4}\,\delta({\rm m_{e}c^2}-E)+3\times\frac{3}{4}\,\frac{1}{N_{3\gamma}}\frac{d N_{3\gamma}}{dE}\big)\,,
\end{split}
\label{annGal}
\end{align}
where $\frac{1}{N_{3\gamma}}\frac{d N_{3\gamma}}{dE}$ is defined by Eq.\,(\ref{f3}) letting $z=0$.

We calculate the photon and positron spectra from a single PBH with \texttt{BlackHawk}.\,2.0~\cite{Arbey:2021mbl}. Fig.\,\ref{fig:EGGal} shows the contributions from the Galactic and extragalactic direct/indirect components to the total simulated IGRB flux in Eq.\,(\ref{Ftot}), for $M=7\times 10^{16}\rm g$, $f=1$ (top panel) and $M=10^{15}\rm g$, $f=1$ (bottom panel) independently. As one can see, the flux (red line) from the Galactic direct HR can be several times larger than the extragalactic one (blue line) at around the peak of the spectrum, especially for massive PBHs. The component from the $e^{+}$ annihilation can be larger than direct one, which would play an important role in contributing to the 100-511\,keV IGRB.

\section{Datasets and the astrophysical background modeling}
\label{secIII}
\begin{figure}[htbp]
  \centering
    \includegraphics[width=0.48\textwidth]{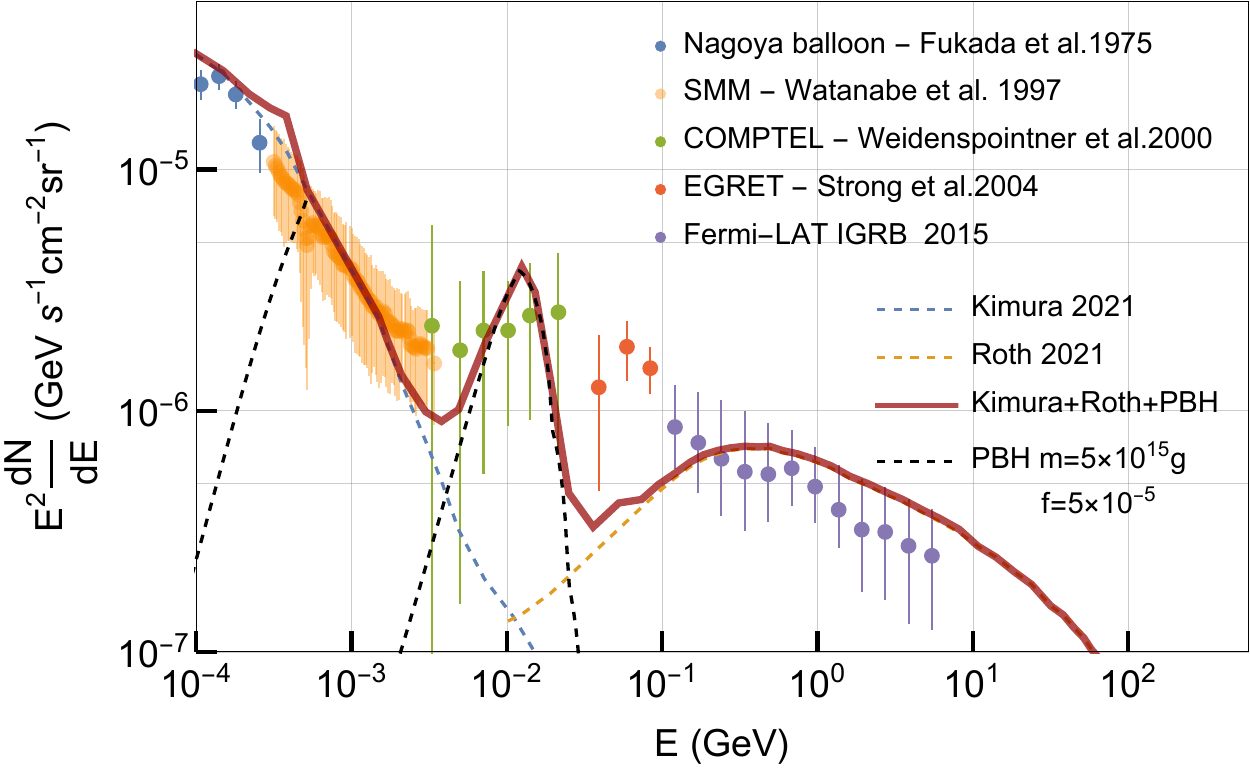}
    \caption{The IGRB spectrum as measured by Fermi-LAT~\cite{Fermi-LAT:2014ryh}, EGRET~\cite{EGRET:1997},
    COMPTEL~\cite{Comptel:1995}, SMM~\cite{SMM:1997} and Nagoya balloon missions~\cite{Fukada:1975}. The Fermi-LAT data corresponds to the foreground model C~\cite{Fermi-LAT:2014ryh}, and the Galactic-foreground modeling uncertainty is added in quadrature to the data with $1 \sigma$ stochastic error. Kimura model~\cite{Kimura:2020thg} (blue dashed line) represents the contribution of low-luminosity AGN to the IGRB flux, Roth model~\cite{Roth:2021lvk} (red dashed line) represents the contribution from star-forming galaxies (SFG), and the PBHs's contribution is represented with dashed black line. The sum of the three models is also shown with the red                                       line.}
    \label{fig:IGRB}
\end{figure}
The IGRB is the measured gamma-ray emission including all unresolved extragalactic emissions in a given survey and any residual (approximate) isotropic Galactic foregrounds~\cite{Fermi-LAT:2014ryh}. In this study, we use the observed IGRB of HEAO-1+balloon, SMM, COMPTEL, EGRET and Fermi-LAT (foreground model C) from 100\,keV to 5\,GeV \footnote{In general, IGRB is referred to as observations of~\cite{Fermi-LAT:2014ryh} and~\cite{EGRET:1997}. For convenience, all data used in this paper is broadly referred to as IGRB.} as shown in Fig.\,\ref{fig:IGRB}, which corresponds to the region where the PBH HR in the mass range $10^{13}$\,g-$10^{18}$\,g is expected to contribute to the IGRB significantly. The Fermi-LAT data with energy $\gtrsim$\,6\,GeV is not chosen because of the two following facts. The background above GeV is expected to be mainly contributed by the $10^{14}$-$10^{15}$\,g PBHs~\cite{Carr:2009jm}. Though the HR of $10^{13}$-$10^{14}$\,g PBHs, whose lifetime is shorter than the present age of the Universe ($t_0\approx\tau(5\times10^{14}\,\rm g)$), is expected to be concentrated at GeV band (the BH temperature $T_{\rm BH}\approx1\,\frac{10^{13}\,\rm g}{M}$\,GeV~\cite{Carr:2009jm,Belotsky2014}) in the co-moving reference frame, it would be redshifted to MeV band in the observed reference frame. Therefore, the HR from our considered PBHs should not significantly contribute to the IGRB above GeV. Secondly, since a considerable part of HR photons for energy $\gtrsim$\,10\,GeV propagating over cosmological distances would be absorbed by soft background (e.g.,\,extragalactic background light) photons via electron-positron pair production, only low-redshift ($z\lesssim8$) PBHs could significantly contribute to the IGRB with corresponding energy.

Considerable efforts have been devoted to interpreting
the IGRB in terms of a superposition of many unresolved extragalactic gamma-ray sources. It is widely believed that the observations between 100 to 200\,keV are predominantly produced by coronal thermal emission from radio-quiet AGN (Seyferts)~\cite{Ueda:2014tma}. Recently, Ref.\,\cite{Kimura:2020thg,Murase:2019vdl} and~\cite{Inoue:2019fil} have explained the MeV (100\,keV to several\,MeV) IGRB together with PeV neutrinos background as the accretion-disk emission in low-luminosity AGN. Meanwhile, other possible candidates including radio-loud AGN~\cite{Inoue:2011bm,Ajello:2012,Toda:2020msp}, Kilonovae and type-Ia supernovae, are found to only contribute a limited share to the MeV IGRB~\cite{Strigari:2005hu,Ruiz-Lapuente:2015yua}. We take into account the simulated IGRB of Kimura\,{\it\,et\,al}~\cite{Kimura:2020thg} into the analysis for PBH constraint, and the model from~\cite{Inoue:2019fil} is also discussed in Sec.\,\ref{sec:discussion}. The former seems more realistic~\cite{Gutierrez:2021vnk,Kheirandish:2021wkm}, where the thermal electrons inside hot accretion flows naturally emit soft gamma-rays via synchrotron self-Compton processes.

For the \emph{100\,MeV-5\,GeV} IGRB, the primary candidate sources provided the bulk of such backgrounds are unresolved SFGs~\cite{Fields:2010,Lacki:2014,Tamborra:2014xia,Xia:2015wka,
Linden:2016fdd,Roth:2021lvk} and radio galaxies (RG, misaligned radio-loud AGN)~\cite{DiMauro:2013xta} (Blazar, aligned radio-loud AGN, should be the primary candidate at energy above $\sim$\,5\,GeV too~\cite{DiMauro:2013zfa,Qu:2019zln}). However, the contribution for SFG given by \cite{DiMauro:2013xta} has huge uncertainty from 10\% to 100\%, and a recent study of~\cite{Stecker:2019ybn} finds that RGs only contribute 4\%-18\% of the IRGB using a large sample of Fermi-LAT RGs. The SFG origin for this IGRB is strongly favored by the recent work of~\cite{Roth:2021lvk}, whose method is based on a physical model for the gamma-ray emission with no free parameters (all quantities that are measured directly) rather than simple empirical scalings. In addition, the statistical analyses of angular fluctuations in the IGRB and cross-correlations between IGRB and galaxies also support the SFG origin~\cite{Xia:2015wka,Ando:2017alx}. Therefore, we model the astrophysical component in the IGRB below 5\,GeV with the SFG contributions provided by~\cite{Roth:2021lvk}.

The red dashed line in Fig.\,\ref{fig:IGRB} represents the Roth\,{\it et\,al} model. The total flux from simulated ``SFGs\,(Roth\,2021)+AGNs\,(Kimura\,2021)+PBHs'' with $M=5\times10^{15}$\,g and $f=8\times10^{-5}$ (black dash line) is represented by the red line. In this scheme, the IGRB round 10\,MeV band still cannot be resolved, and thus we can expect that it gives a relatively weak constraint on the PBHs abundance in the relevant mass range.

\section{Results: constraints on the PBH abundance}
\label{sec:result}
In this section, we will present our results about the constraint on the
PBH abundance $f$ with monochromatic mass distribution in the interested mass range. Since angular momentum makes PBHs evaporate faster and thus makes the bounds on
their abundance stronger, we assume a population of non-rotating PBHs in our simulation~\cite{Ballesteros:2019exr}. In this sense, our results are conservative compared to the case considering rotating PBHs.
\subsection{Conservative constraint: without astrophysical component modeling}
\begin{figure}[htbp]
  \centering
    \includegraphics[width=0.45\textwidth]{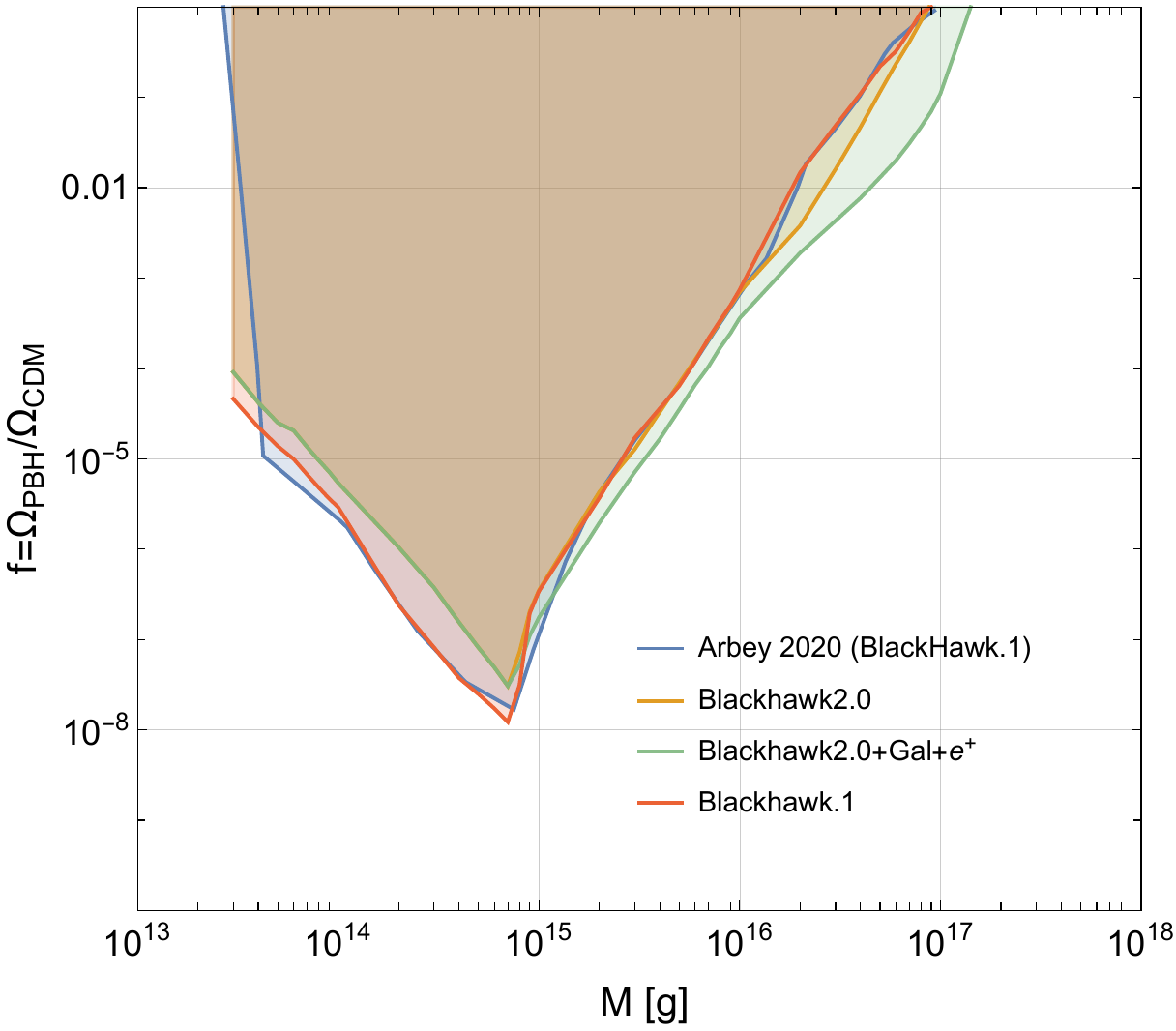}
    \caption{68\% C.\,L. bounds derived without astrophysical background modeling. The green line represents the bound considered the direct Galactic PBHs HR and the radiation due to $e^{+}$ annihilation, where the secondary spectra (and thus the whole HR flux) is reliably simulated with \texttt{BlackHawk}.\,2.0 or Eq.\,(\ref{FSR}). The orange line represents the bound without contribution from the Galactic PBHs and $e^{+}$, where the PBH flux is obtained with \texttt{BlackHawk}.\,2.0 too. For comparison, we also show the constraint in the literature~\cite{Arbey:2019vqx} by blue line and our bound (by red line) where we calculate the HR with the way of this literature.}
    \label{fig:PBHf1}
\end{figure}
We derive conservative bounds on the PBH abundance without above astrophysical component modeling.
These constraints require that the flux from PBHs, Eq.\,(\ref{Ftot}), does not exceed any measured IGRB data-points by more than 1\,$\sigma$, as done in~\cite{Ballesteros:2019exr}. The bound thus obtained is displayed in Fig.~\ref{fig:PBHf1} (green line), compared with that in the literature~\cite{Arbey:2019vqx} (blue line). We rule out the totality of DM in the form of PBHs for $10^{13}$\,g\,$\lesssim M\lesssim 10^{17}\,$g with 68\,\% C.\,L.
The improvement over the upper limit given by~\cite{Arbey:2019vqx} is significant, especially more than an order of magnitude improvement at $2\times10^{16}$\,g\,$\lesssim M\lesssim 10^{17}\,$g.

In order to distinguish where the improvement comes from, we also show the result (orange line) without the Galactic component of PBH HR and the indirect $e^{+}$-annihilation component in Fig.\,\ref{fig:PBHf1}. As we can see, the improvement at $3\times10^{13}$\,g\,$\lesssim M\lesssim 7\times10^{14}$\,g results from the reliable calculation for the secondary spectra with \texttt{BlackHawk}.\,2.0 or Eq.\,(\ref{FSR}). As the photon flux calculated with Eq.\,(\ref{FSR}) become stronger at relevant energy than that with \texttt{BlackHawk}.\,1 (see the bottom panel in Fig.\,\ref{fig:FSR}) for lighter PBHs, our constraint turns from more stringent to weaker at $M\lesssim7\times10^{14}$\,g. At $M\lesssim 10^{15}$\,g, the additional component to the flux leads to the improvement. The reliable secondary-spectra calculation effects the improvement around $3\times10^{16}$\,g too.

 We can repeat the bound of~\cite{Arbey:2019vqx} well, which verifies our calculations, see the red line in Fig.~\ref{fig:PBHf1}.
\subsection{With astrophysical component modeling}
\label{sec:result_B}

\begin{figure}[htbp]
  \centering
    \includegraphics[width=0.45\textwidth]{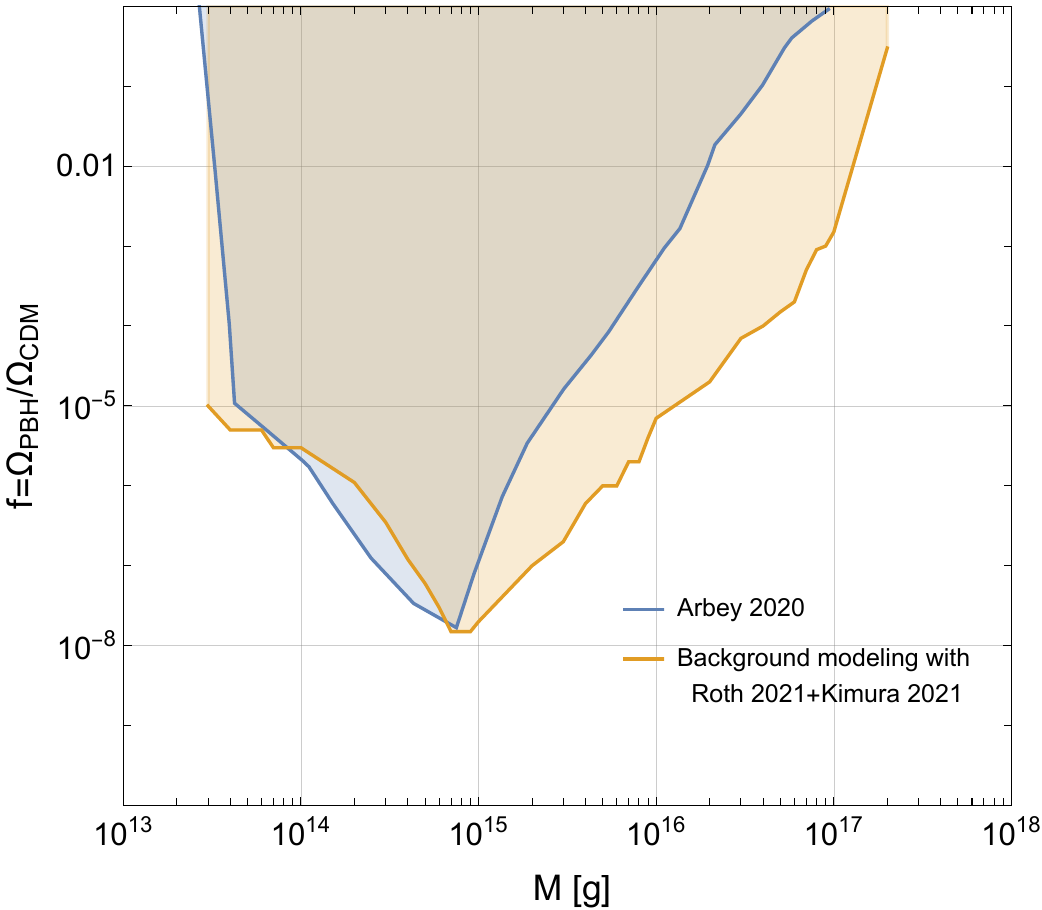}
    \caption{Bound derived with astrophysical background modeling. The orange line represents the upper limit with 95\% C.L. counting for the AGNs and SFGs contribution given by~\cite{Kimura:2020thg} and~\cite{Roth:2021lvk}, respectively (see, Fig.\,\ref{fig:IGRB}). For comparison, we also show the constraint with 68\% C.L. in the literature~\cite{Arbey:2019vqx} by blue line.}
    \label{fig:f_Kimura}
\end{figure}
 Since the measured IGRB flux should contain the components from gamma-ray sources, such as AGNs and SFGs, more strength and realistic bounds on $f$ would be obtained if the components are accounted. Therefore, we add the astrophysical component given by~\cite{Kimura:2020thg} and ~\cite{Roth:2021lvk} (see the ``Kimura\,2021'' and ``Roth\,2021'' lines in Fig.\,\ref{fig:IGRB}) into our simulated flux from the PBHs. Since the model~\cite{Roth:2021lvk} is beyond the upper limit of the data-points at several GeVs, the uncertainties in Galactic-foreground modeling are added in quadrature to the Fermi-LAT data\footnote{If the model is still beyond the upper limit of the data-points, the statistical uncertainty is expanded to several $\sigma$ so that the model value is below the upper limit of the data.}~\cite{Fermi-LAT:2014ryh}, and thus, our constraint is derived if the simulated flux does not exceed any measured IGRB data-points by more than 2\,$\sigma$.

Fig.\,\ref{fig:f_Kimura} shows our bound with orange line and that in~\cite{Arbey:2019vqx} (with 68\% C.L.) for comparison. Our improvements in simulating the diffuse flux, i.e.,\,calculating the secondary spectra with \texttt{BlackHawk}.\,2.0 and including the $e^{+}$-annihilation radiation, Galactic PBH component, and the astrophysical background, lead to more than an order of magnitude tighter constraint at $10^{15}$\,g\,$\lesssim M\lesssim 10^{17}\,$g, pushing the constrained PBH mass to $M=2\times 10^{17}\,$g. Since the ``Kimura+Roth'' model can only account for a small fraction of the IGRB at 20-100\,MeV where the HR emitted by PBHs of $10^{14}$\,g\,$\lesssim M\lesssim 10^{15}\,$g is concentrated, the improvement at this mass band due to the inclusion of the background is not significant.

%background from SFGs can explain all the measured IGRB at several GeV. Conversely, the constraint at around $10^15$\,g is weaker since the ``Kimura+Roth'' model can only account for a small fraction of the IGRB at 20-100\,MeV.
%%Compare to the significant divergence between our bound and Arbey2021 at small mass in Fig.\,\ref{fig:PBHf1}, the divergence narrows as the addition of background offset the reduction of the secondary spectra from the correction.

\section{Discussion}
\label{sec:discussion}
\begin{figure}[htbp]
  \centering
    \includegraphics[width=0.47\textwidth]{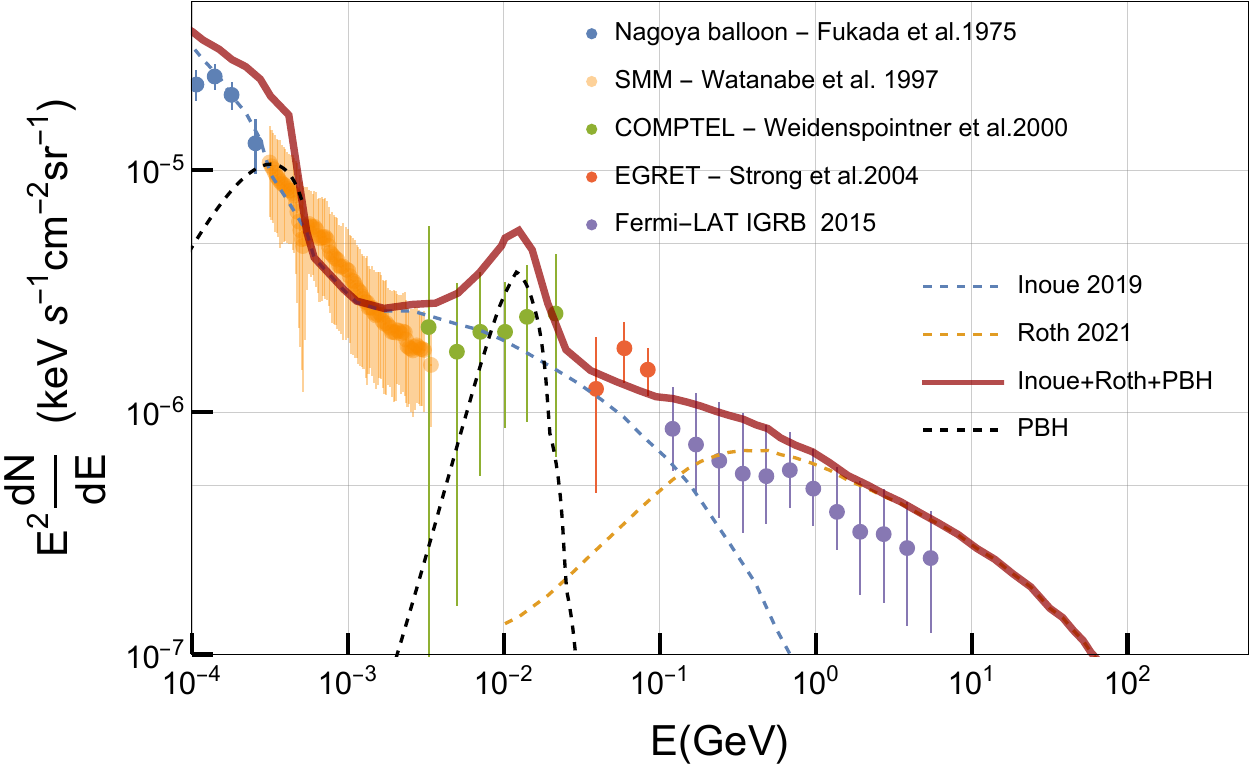}\\

    \includegraphics[width=0.45\textwidth]{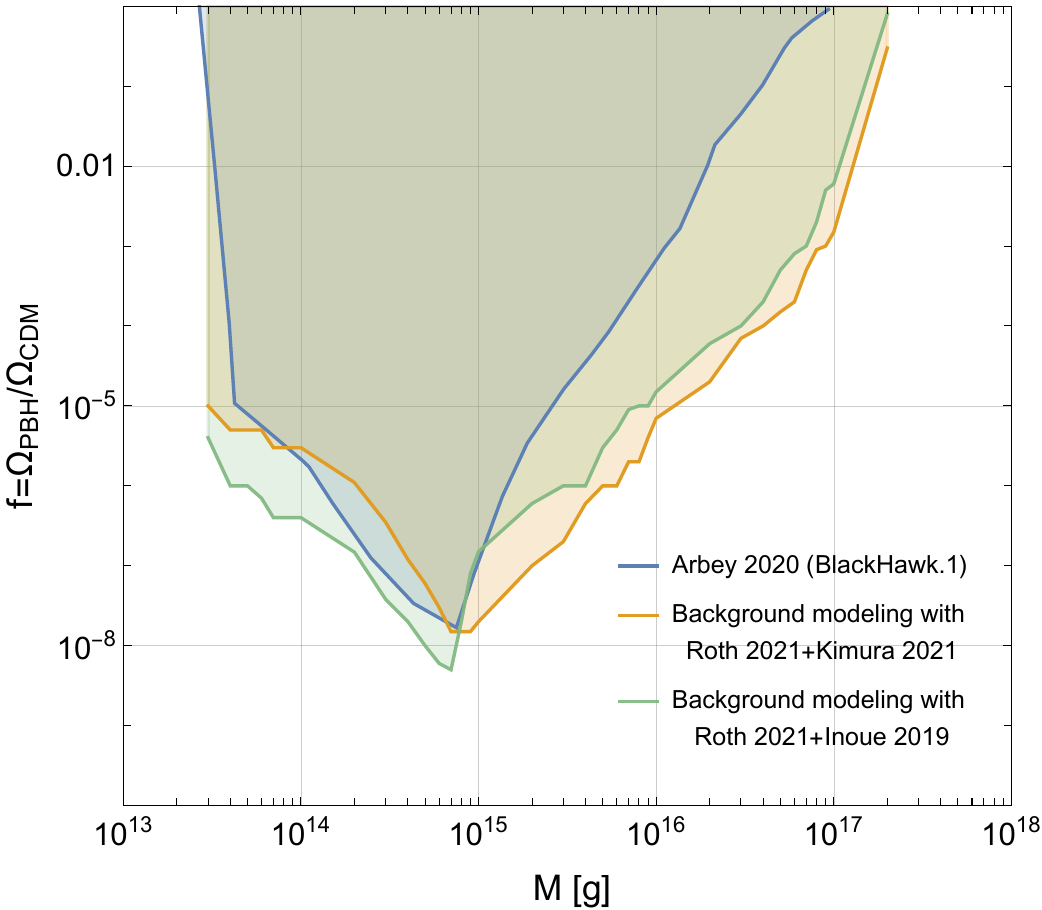}
    \caption{Top panel: Same as Fig.\,\ref{fig:IGRB}, but the MeV astrophysical background is considered with the model in~\cite{Inoue:2019fil} (blue dashed line). The sum of models Inoue\,2019~\cite{Inoue:2019fil}, Roth\,2021~\cite{Roth:2021lvk} and PBH is shown with maroon line. Bottom panel: bounds at
    95\%\,C.L. accounting for astrophysical background modeling. The green line represents the result of the AGNs and SFGs contribution given by~\cite{Inoue:2019fil} and~\cite{Roth:2021lvk}, respectively. For comparison, we also show the constraint in the literature~\cite{Arbey:2019vqx} by blue line and that for ``Roth\,2021+Kimura\,2021" model in Fig.\,\ref{fig:f_Kimura}.}
    \label{fig:IGRB2}
\end{figure}
In this section, we will discuss other model explaining the measured IGRB.

The model in~\cite{Inoue:2019fil} can also explain the MeV background together with PeV neutrinos background with the accretion-disk emission in low-luminosity AGN. Based
on the observation evidence of nonthermal synchrotron emission in two nearby Seyfert galaxies, the MeV gamma-rays are generated by nonthermal electrons via inverse Compton scattering of disk
photons in this model, rather than thermal electrons via Comptonization of their synchrotron photons as in~\cite{Kimura:2020thg}. Fig.\,\ref{fig:IGRB2} (top) displays this model with blue dash line and the total of the three models (Inoue\,2019~\cite{Inoue:2019fil}+Roth\,2021~\cite{Roth:2021lvk}+PBHs) with maroon line.

The derived bound using the analysis method described in Sec.~\ref{sec:result_B} is reported in the bottom panel of Fig.\,\ref{fig:IGRB2} by green line. Not surprisingly, the advance in the treatment of the simulated flux notably improves the constraint on $f$ over that from~\cite{Arbey:2019vqx} (blue line) at most of the considered mass range.
The ``Inoue\,2019 + Roth\,2021" bound is more conservative at high mass band $10^{15}$\,g\,$\lesssim M\lesssim 2\times10^{17}\,$g but more stringent below $10^{15}$\,g than the ``Kimura\,2021 + Roth\,2021" one (green line). These phenomena can be attributed to the relative size of the contributions of the two background models to the IGRB. Namely, the flux modeled by Kimura\,2021 (blue dashed line in the top panel of Fig.\,\ref{fig:IGRB}) above 2\,MeV is more conservative than Inoue\,2019~\cite{Inoue:2019fil} but slightly higher at 100\,keV-2\,MeV. These results suggest that the Fermi-LAT observations only mainly affect the constraint on the PBHs around $8\times10^{14}$\,g.

Future MeV space missions, such as GRAMS~\cite{Aramaki2020}, AMEGO~\cite{McEnery2019,Ray:2021mxu}, XGIS-THESEUS~\cite{Ghosh2021}, and e-ASTROGAM~\cite{Angelis2017}, will be able to detect a larger number of point sources and will help to verify the models for the unresolved astrophysical background and thus test our constraints on the PBHs.

%The small differences at $3\times10^{15}$\,g\,$\lesssim M\lesssim 5\times10^{16}\,$g between our results for different background modeling, mean that our result is robust at this mass range.

\section{Conclusion}
\label{sec:conclusion}
In this article, a population of PBHs with monochromatic mass distribution in the range $10^{13}\,{\rm g} - 10^{18}\,$g hase been considered. New upper limits on the PBH abundance $f$ constituting all or part of the DM have been set by comparing the measured IGRB flux at 100\,keV-5\,GeV band with the diffuse flux from the Galactic and extragalactic Hawking radiation.

  We have reliably calculated the secondary spectra produced by primary particles for energy below 5\,GeV according to Eq.\,(\ref{fig:FSR}), with the latest public code \texttt{BlackHawk}.\,2.0~\cite{Arbey:2021mbl}. As a result, the conservative upper limit at $3\times10^{13}$\,g\,$\lesssim M\lesssim 7\times10^{14}\,$g is several times weaker than the relevant one of Arber\,{\it et\,al.}~\cite{Arbey:2019vqx} but stronger at around $3\times10^{16}\,$g, see Fig.\,\ref{fig:PBHf1}. Furthermore, our model takes into account the Galactic PBHs' diffuse radiation and that of the positronium annihilation from evaporated positrons, see Eq.\,(\ref{annEG}) and\,(\ref{EqGal}). The resulting constraint on $f$ is significantly improved over literature Arber\,{\it et\,al}.~\cite{Arbey:2019vqx}, especially more than an order of magnitude at $2\times10^{16}$\,g\,$\lesssim M\lesssim \times10^{17}\,$g, see Fig.\,\ref{fig:PBHf1}.

  After further accounting for the astrophysical background model given by~\cite{Kimura:2020thg,Murase:2019vdl} and~\cite{Roth:2021lvk}, the constraint is tighter more than an order of magnitude at $10^{15}$\,g\,$\lesssim M\lesssim 10^{17}\,$g compared to~\cite{Arbey:2019vqx}, pushing the constrained PBH mass to $M=2\times 10^{17}\,$g, see Fig.\,\ref{fig:f_Kimura}.

   Other constraints, based on the alternative astrophysical model~\cite{Inoue:2019fil} explaining the MeV IGRB, are also discussed (see Fig.\,\ref{fig:IGRB2}). They are helpful to comprehend the uncertainties of our results from the astrophysical background modeling. For general extended mass function, the corresponding bounds should be tighter or similar (see e.g.,~\cite{Iguaz:2021irx,Arbey:2019vqx}). Future MeV telescopes are hoped to test our constraints with background modeling.

\begin{acknowledgements}
We thank Seishi\,Enomoto, Yi-Lei\,Tang, and Chengfeng\,Cai for useful discussions and comments.
This work is supported by the National Natural Science Foundation of China (NSFC) under Grant No. 11875327, the Fundamental Research Funds for the Central Universities, China, and the Sun Yat-sen University Science Foundation.
\end{acknowledgements}

\end{document}